\documentclass[preprint, sort&compress, 3p]{elsarticle}

\usepackage{lineno,hyperref}

\journal{IMAC-XLII conference}

\graphicspath{{figures/}}
\usepackage{amsmath,amsfonts,amssymb}
\usepackage{bm}
\DeclareMathOperator*{\argmax}{arg\,max}

\usepackage[skip=0pt]{caption}
\usepackage[inline]{enumitem}
\usepackage{subcaption}
\usepackage{multicol}
\usepackage{xcolor}
\usepackage[normalem]{ulem}
\usepackage[nameinlink,capitalize]{cleveref}
\usepackage{booktabs}

\begin{document}
	
	\begin{frontmatter}
		
		\title{Bayesian decision-theoretic model selection for monitored systems}
		
		\author[mysecondaryaddress]{Antonios Kamariotis \corref{mycorrespondingauthor}}
		\cortext[mycorrespondingauthor]{Corresponding author}
		\ead{antonios.kamariotis@ibk.baug.ethz.ch}
        \author[mysecondaryaddress]{Eleni Chatzi}
		\ead{chatzi@ibk.baug.ethz.ch}
		
		\address[myprimaryaddress]{Institute of Structural Engineering, ETH Zurich, Stefano-Franscini-Platz 5, 8093 Zurich, Switzerland}
  
        \begin{abstract}
	Engineers are often faced with the decision to select the most appropriate model for simulating the behavior of engineered systems, among a candidate set of models. Experimental monitoring data can generate significant value by supporting engineers toward such decisions. Such data can be leveraged within a Bayesian model updating process, enabling the uncertainty-aware calibration of any candidate model. The model selection task can subsequently be cast into a problem of decision-making under uncertainty, where one seeks to select the model that yields an optimal balance between the reward associated with model precision, in terms of recovering target Quantities of Interest (QoI), and the cost of each model, in terms of complexity and compute time. In this work, we examine the model selection task by means of Bayesian decision theory, under the prism of availability of models of various refinements, and thus varying levels of fidelity. In doing so, we offer an exemplary application of this framework on the IMAC-MVUQ Round-Robin Challenge. Numerical investigations show various outcomes of model selection depending on the target QoI.
        \end{abstract}
		
		
	\end{frontmatter}

\section{Introduction}

    The task of selecting the optimal model for simulating the behavior of engineered systems from a pool of candidate models is a common and critical challenge in engineering \cite{Beck_2004,YIN2019306}. The candidate set of possible models may range, e.g., from the simplest analytical model to a high-fidelity finite element model. The model selection problem, which is characterized by significant uncertainty, fundamentally underpins the reliability and efficiency of system design and operation. In this context, experimental monitoring data emerge as a valuable asset, reducing the uncertainty and offering substantial support to engineers in their decision-making process \cite{Straub_2017,HUGHES2022108569,KAMARIOTIS2023109708}. Conditional on availability of such data, any available model can be updated/calibrated via a Bayesian model updating (BMU) process \cite{BEHMANESH2015360,KAMARIOTIS2022108465}. This process leads to the estimation of the posterior uncertainty associated with each model, thereby facilitating a data-informed decision-making process. The model selection task, conditioned on a given set of observations, turns into a problem of decision-making under uncertainty (DMUU) \cite{Berger_1985}. The objective is to strike a balance between the precision of a model, in terms of recovering QoIs derived from the experimental observations, and the associated cost. This paper delves into the model selection task through the lens of Bayesian decision theory \cite{Berger_1985}, which serves for the solution to DMUU problems. We first briefly summarize the Bayesian decision-theoretic approach to model selection. Subsequently, we numerically investigate this approach on the IMAC-MVUQ Round-Robin Challenge \cite{Platz_2023_RR_IMAC}.

	\section{A Bayesian decision-theoretic approach to model selection}
    The problem of selecting the appropriate engineering model, given a set of observations, can be cast as a DMUU problem, whereby the optimal decision is to employ the model that maximizes the expected utility \cite{Raiffa_1961}. In decision theory, utility provides the formal mathematical context that assesses the optimality of different decision alternatives. A utility function, whose form is defined by the decision maker, is a mathematical function that assigns a real number, the utility, to the values of the attributes of a decision problem. For the problem of selecting the best model to use, two main attributes can be identified:
        \begin{enumerate}
        \item model precision, in terms of recovering target QoIs, and 
        \item the cost of each model, in terms of computational complexity \cite{Arora_2006} and runtime.   
        \end{enumerate}
        These attributes are potentially competing, as one could reasonably expect that the higher the precision of a model, the higher its cost.
	
	Bayesian decision theory \cite{Berger_1985} offers a mathematical framework for the solution to DMUU problems, in which data of some sort become available and assist in reducing the uncertainty in the problem.	A random vector $\boldsymbol{\theta}$ is defined, which contains the uncertain parameters of a model. A prior probabilistic model $\pi_{\mathrm{pr}}(\boldsymbol{\theta})$ is typically assigned by the analyst. Conditional on the availability of experimental data $\mathbf{z}$, BMU can be performed, which outputs the posterior distribution of $\boldsymbol{\theta}$ given $\mathbf{z}$, denoted by $\pi_{\mathrm{pos}}(\boldsymbol{\theta}|\mathbf{z})$. The optimal action $a_{\mathrm{opt}}$ to take conditional on data $\mathbf{z}$ is then obtained following Equation~\ref{eq:MEU}:
	\begin{equation}
		a_{\mathrm{opt}|\mathbf{z}} = \argmax_a
		 \mathbb{E}_{\boldsymbol{\theta}|\mathbf{z}}[U(a,\boldsymbol{\theta})],
		\label{eq:MEU}
	\end{equation}
	where $U(a,\boldsymbol{\theta})$ is the utility function, which is a function of the uncertain parameter vector $\boldsymbol{\theta}$ and some decision $a$, and maps possible outcomes to their utility (e.g., see Fig.~\ref{fig:utility_nRMSE}). The expectation in Equation~\ref{eq:MEU} is taken with respect to the posterior distribution $\pi_{\mathrm{pos}}(\boldsymbol{\theta}|\mathbf{z})$. With regard to the model selection task, the possible decisions can be summarized in a set $\{a_1, a_2, \dots, a_n\}$ in the case of $n$ competing models, where $a_i$ represents the decision to employ the $i$-th model for simulating the behavior of an engineered system. 

 	\section{Numerical investigation: IMAC-MVUQ Round-Robin challenge}

	A Round-Robin challenge within IMAC's Model Validation and Uncertainty Quantification (MVUQ) community was introduced in \cite{Platz_2023_RR_IMAC}. One of the challenge's objectives, which we address in this paper, relates to selecting, under the premise of uncertainty, the appropriate model that best simulates the behavior of a given realized dynamical system. Epxerimental measurements (monitoring data) obtained through different instantiations of the system in an experimental test environment are available.

 	\begin{figure}
		\centering
		\begin{subfigure}{.35\textwidth}
			\includegraphics[width=1.\linewidth]{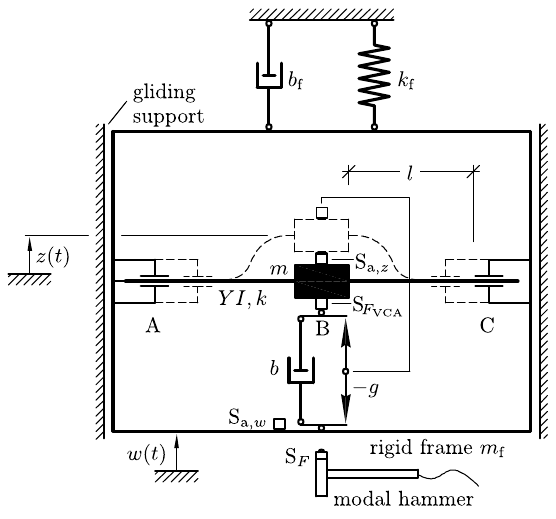}
			\caption{Schematic diagram of experimental setup. Figure taken from \cite{Platz_2023_RR_IMAC}.}
			\label{subfig:experiment}
		\end{subfigure}\hspace{1cm}
		\begin{subfigure}{.18\textwidth}
		\includegraphics[width=1.\linewidth]{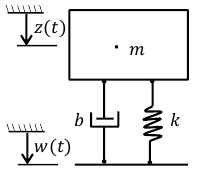}
			\caption{Model 1: one-mass oscillator with base point displacement excitation.}
			\label{subfig:1DOF}
		\end{subfigure}\hspace{1cm}
		\begin{subfigure}{.21\textwidth}
            \includegraphics[width=1.\linewidth]{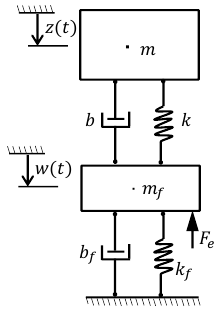}
			\caption{Model 2: two-mass oscillator with impulse force excitation.}
			\label{subfig:2DOF}
		\end{subfigure}
            \vspace{5pt}
		\caption{Dynamical system realization as an experimental test environment and two candidate mathematical models for simulating the system's behavior.}
		\label{fig:analytical_models_experiment}
	\end{figure}
	
	Fig.~\ref{subfig:experiment} schematically represents the experimental test setup, where a one-mass oscillator of mass $m$ is embedded in a frame of mass $m_f>>m$ (for details the reader is referred to \cite{Platz_2023_RR_IMAC}). The frame is excited by a modal hammer impulse-like force. Two acceleration sensors are deployed, one on the oscillating mass $m$, and one on the frame mass $m_f$, measuring the respective absolute accelerations $\ddot{z}(t)$ and $\ddot{w}(t)$, while a force sensor is further deployed to measure the modal hammer impulse force $F_h(t)$. Hence, the set of available observations can be summarized in the measurement vector $\mathbf{z}=[\ddot{z}(t), \ddot{w}(t), F_h(t)]$. 
	
	Fig.~\ref{subfig:1DOF} represents a one-mass oscillator model, which is subjected to a base point displacement excitation. There are three uncertain parameters for this model summarized in the vector $\boldsymbol{\theta}_1=[m,b,k]$. Fig.~\ref{subfig:2DOF} represents an alternative model, namely a two-mass oscillator model with an impulse force $F_e$ excitation, which aims to better represent the realized dynamical system's behavior. This mathematical model contains six uncertain parameters in $\boldsymbol{\theta}_2=[m,b,k, m_f, b_f, k_f]$. For both mathematical models, analytical solutions exist for the vibration response in the frequency domain \cite{Platz_2023_RR_IMAC}. The goal in this section is to decide which of the two candidate analytical models leads to a better balance between model precision and the cost of each model, given the set of experimental monitoring data that is assumed available $\mathbf{z}$. The definition of model precision varies depending on the Quantity of Interest (QoI) that one targets to recover. In the following subsections we consider two different QoIs based on which we quantify model precision.
	
	\subsection*{Model selection with respect to recovering the response amplitude of mass $m$}
	
	Let us assume that the QoI that we wish to recover is the response amplitude of mass $m$. We perform the Fast Fourier Transform (FFT) on the experimental acceleration time series vector $\ddot{z}(t)$ and store the absolute value of the output array corresponding to frequencies $f=0-100$ Hz in a vector $\tilde{\boldsymbol{y}}_{a_z}\in \mathbb{R}^{n_f}$, which is plotted in black in both subfigures of Fig.~\ref{fig:BMU}. For both the one-mass (model 1) and two-mass (model 2) oscillator models, analytical equations for the Transfer Functions (TF) exist, which represent the response amplitude of mass $m$ as a function of the frequency of the input signal \cite{Chopra_1995}. The TF of model 1 is denoted by $V_1(f, \boldsymbol{\theta}_1)$ and is a function of frequency and the random vector $\boldsymbol{\theta}_1$, whereas the TF of model 2 is denoted by $V_2(f, \boldsymbol{\theta}_2)$ and is a function of frequency and the random vector $\boldsymbol{\theta}_2$. The input to model 1 is the base point displacement excitation $w(t)$, which is not measured. However, the experimental dataset contains the measurement of the frame acceleration time series $\ddot{w}(t)$, on which we perform the FFT and obtain the vector $\tilde{\boldsymbol{y}}_{a_w}\in \mathbb{R}^{n_f}$. We then approximate the FT response amplitude of mass $m$ via the TF of model 1 as $V_1(f, \boldsymbol{\theta}_1)\cdot\tilde{\boldsymbol{y}}_{a_w}$. The input to model 2 is the impulse force excitation $F_e$. The experimental data contain the measurement of the modal hammer impulse force time series $F_h(t)$. We perform the FFT on $F_h(t)$, through which we obtain the vector $\tilde{\boldsymbol{y}}_{F_h}\in \mathbb{R}^{n_f}$. Model 2 approximates the FT response amplitude of mass $m$ as \smash{$V_2(f, \boldsymbol{\theta}_2)\cdot \omega^2 \cdot \tilde{\boldsymbol{y}}_{F_h}$, where $\omega=2\pi f$}.
	
	\subsubsection*{Bayesian model updating}
	
	\begin{table}
            \small
		\centering
		\caption{Prior probabilistic model for the uncertain parameters of models 1,2 chosen based on \cite{Lenz_2020}.}
		\label{table:prior}
		\begin{tabular}{@{}lcccccc@{}}
			\toprule
			& $k$ & $m$ & $D$ & $k_f$ & $m_f$ & $D_f$\\ \midrule
			Distribution & Normal & Normal & Normal & Normal & Normal & Normal\\
			mean & 38494 & 0.925 & 0.12 & 722 & 9.33 & 0.03\\
			c.o.v. & 6.6\% & 5.3\% & 10\% & 6.6\% & 10\% & 15\%\\ \bottomrule
		\end{tabular}
	\end{table}

\begin{figure}
		\centering
		\begin{subfigure}{.485\textwidth}
			\includegraphics[width=1.\linewidth]{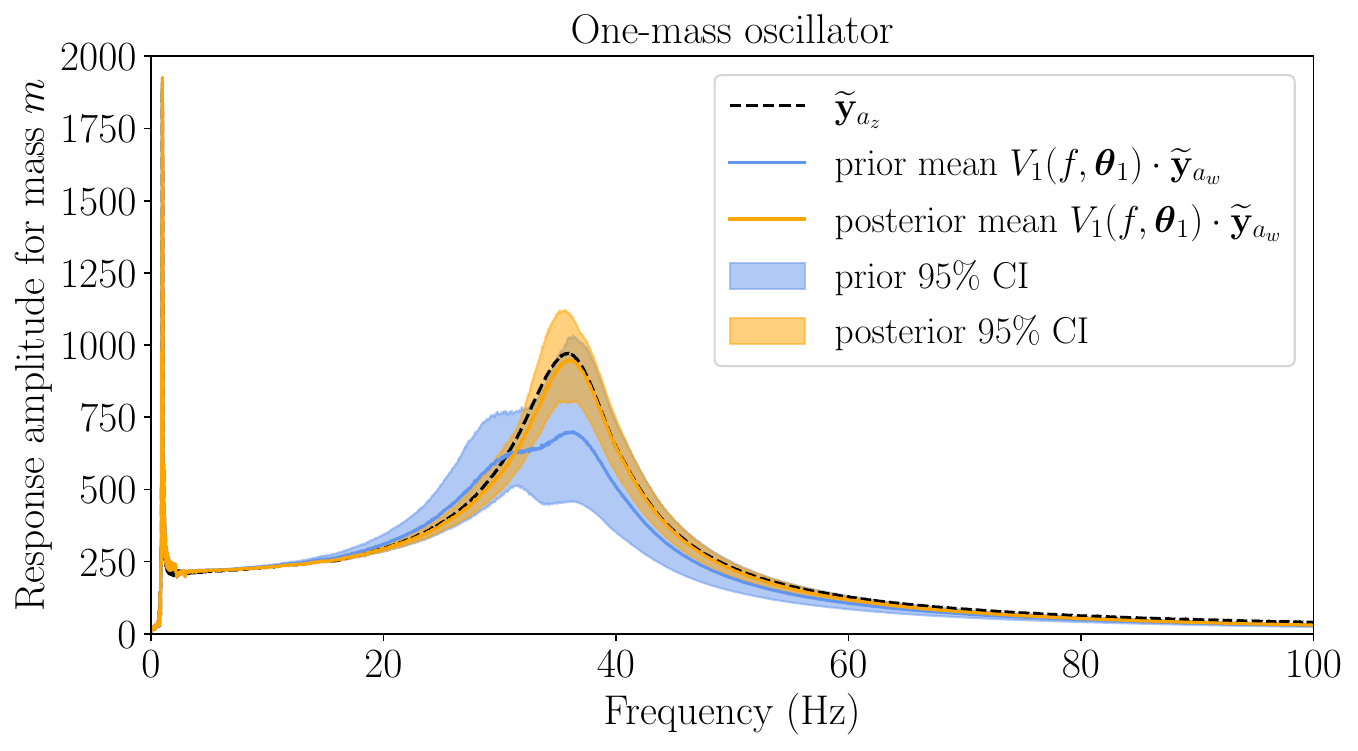}
			\caption{Prior and posterior curves for the response amplitude of mass $m$ obtained with model 1.}
			\label{subfig:1DOF_updating}
		\end{subfigure}
		\begin{subfigure}{.485\textwidth}
			\includegraphics[width=1.\linewidth]{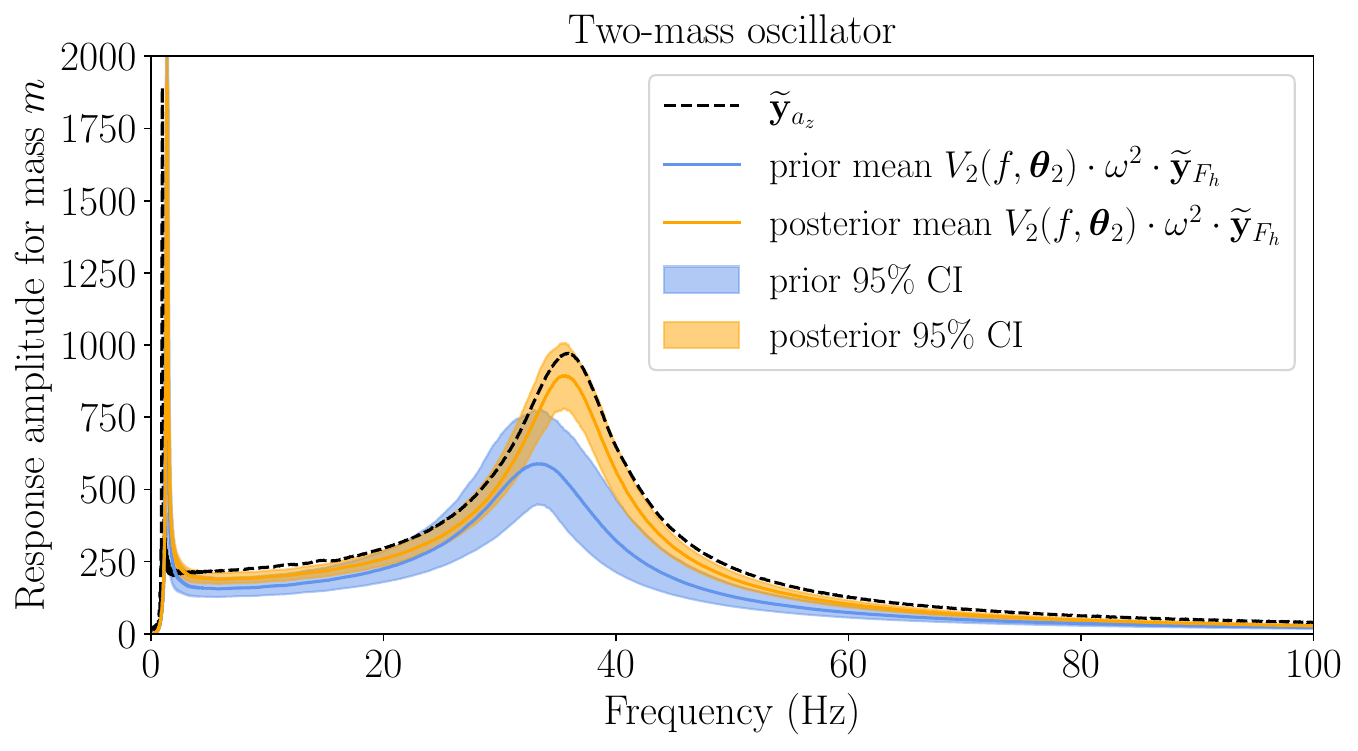}
			\caption{Prior and posterior curves for the response amplitude of mass $m$ obtained with model 2.}
			\label{subfig:2DOF_updating}
		\end{subfigure}
            \vspace{5pt}
		\caption{Bayesian model updating (BMU) for obtaining the posterior distribution of the uncertain model parameters and subsequently the posterior curves for the model-predicted response amplitude of mass $m$.}
		\label{fig:BMU}
	\end{figure}

	The prior probabilistic models assigned for the uncertain parameters of models 1,2 contained in $\boldsymbol{\theta}_1$, $\boldsymbol{\theta}_2$ are summarized in Table~\ref{table:prior}. \smash{$D=bm/2k^2$} and \smash{$D_f=b_fm_f/2k_f^2$} are the damping ratios. We draw $n_s=$1000 samples from these prior distributions and in Figs.~\ref{subfig:1DOF_updating}, ~\ref{subfig:2DOF_updating} we plot in blue the prior mean estimate and 95\% credible intervals (CI) for $V_1(f, \boldsymbol{\theta}_1)\cdot\tilde{\boldsymbol{y}}_{a_w}$ and \smash{$V_2(f, \boldsymbol{\theta}_2)\cdot \omega^2 \cdot \tilde{\boldsymbol{y}}_{F_h}$} that are obtained with these samples. Conditional on the experimental vector $\tilde{\boldsymbol{y}}_{a_z}$, a BMU step can be performed for obtaining the posterior distribution of the uncertain model parameters. We assume the following probabilistic model for the discrepancy between the experimental and the model-predicted FT response amplitude for mass $m$:
	\begin{equation}
		\text{model 1:}\,\,\,\,\,\,\boldsymbol{\eta}_1= \tilde{\boldsymbol{y}}_{a_z} - V_1(f, \boldsymbol{\theta}_1)\cdot\tilde{\boldsymbol{y}}_{a_w} \in \mathbb{R}^{n_f} \sim \mathcal{N}\big(\boldsymbol{0}, \boldsymbol{\Sigma}=\mathrm{diag}(c^2||\tilde{\boldsymbol{y}}_{a_z}||^2)\big),
	\end{equation}
	\begin{equation}
	\text{model 2:}\,\,\,\,\,\,\boldsymbol{\eta}_2= \tilde{\boldsymbol{y}}_{a_z} - V_2(f, \boldsymbol{\theta}_2)\cdot \omega^2 \cdot \tilde{\boldsymbol{y}}_{F_h} \in \mathbb{R}^{n_f} \sim \mathcal{N}\big(\boldsymbol{0}, \boldsymbol{\Sigma}=\mathrm{diag}(c^2||\tilde{\boldsymbol{y}}_{a_z}||^2)\big),
	\end{equation}
	where $\mathcal{N}(\boldsymbol{0},\boldsymbol{\Sigma})$ denotes the multivariate normal (MVN) distribution with a zero-mean vector and covariance matrix $\boldsymbol{\Sigma}$. A diagonal covariance matrix is assumed, with the variance of each component in the vectors $\boldsymbol{\eta}_1$, $\boldsymbol{\eta}_2$ assumed proportional to the $L_2$-norm of $\tilde{\boldsymbol{y}}_{a_z}$. The factor $c$  can be regarded as a coefficient of variation (c.o.v.), and its chosen value reflects the total prediction error. It is here assumed that $c=$0.05. The likelihood functions for models 1, 2 are written as:
	\begin{equation}
		\text{model 1:}\,\,\,\,\,\,L(\boldsymbol{\theta}_1; \tilde{\boldsymbol{y}}_{a_z}) \sim \mathcal{N}\big(\boldsymbol{\eta}_1; \boldsymbol{0},\boldsymbol{\Sigma}\big),
	\end{equation}
	\begin{equation}
		\text{model 2:}\,\,\,\,\,\,L(\boldsymbol{\theta}_1; \tilde{\boldsymbol{y}}_{a_z})  \sim \mathcal{N}\big(\boldsymbol{\eta}_2;\boldsymbol{0}, \boldsymbol{\Sigma}\big),
	\end{equation}
	where $\mathcal{N}(\cdot \, ;\boldsymbol{0},\boldsymbol{\Sigma})$ denotes the value of the MVN density function at a specified location. For both models 1 and 2, we perform the BMU process by employing the TMCMC algorithm \cite{Ching_2007}, thereby obtaining $n_s=1000$ samples from the posterior distributions \smash{$\pi_{\mathrm{pos}}(\boldsymbol{\theta}_1|\tilde{\boldsymbol{y}}_{a_z})$ and $\pi_{\mathrm{pos}}(\boldsymbol{\theta}_2|\tilde{\boldsymbol{y}}_{a_z})$}. With these posterior samples we compute the posterior mean and 95\% CIs for \smash{$V_1(f, \boldsymbol{\theta}_1)\cdot\tilde{\boldsymbol{y}}_{a_w}$ and $V_2(f, \boldsymbol{\theta}_2)\cdot \omega^2 \cdot \tilde{\boldsymbol{y}}_{F_h}$}, which we plot in orange in Figs.~\ref{subfig:1DOF_updating}, ~\ref{subfig:2DOF_updating}. In both subfigures, one can clearly identify the benefit of exploiting the experimental measurements via the BMU process. For both models 1 and 2, the posterior curves for the model-predicted response amplitude for mass $m$ show significant improvement in capturing the experimental quantity $\tilde{\boldsymbol{y}}_{a_z}$ compared to the prior ones, with a pronounced reduction of the uncertainty in the posterior predictions.
	
	It should be noted that, for both models 1 and 2, two peaks are observed in the model-predicted FT response amplitude plots. The first peak at lower frequencies corresponds to the first eigenfrequency of the frame mass $m_f$, whereas the second peak corresponds to the first eigenfrequency of the oscillating mass $m$. In principle, the one-mass oscillator model 1 should not be able to capture both peaks. However, as shown in Fig.~\ref{subfig:1DOF_updating}, model 1 does capture both peaks. This is only because we here claim to have access to the measured base acceleration input $\ddot{w}(t)$, and its Fourier transformed vector $\tilde{\boldsymbol{y}}_{a_w}$, which contains this first peak. Hence this peak appears also in $V_1(f, \boldsymbol{\theta}_1)\cdot\tilde{\boldsymbol{y}}_{a_w}$.
	
	\subsubsection*{Model selection}
	
\begin{figure}
		\centering
		\includegraphics[width=.40\linewidth]{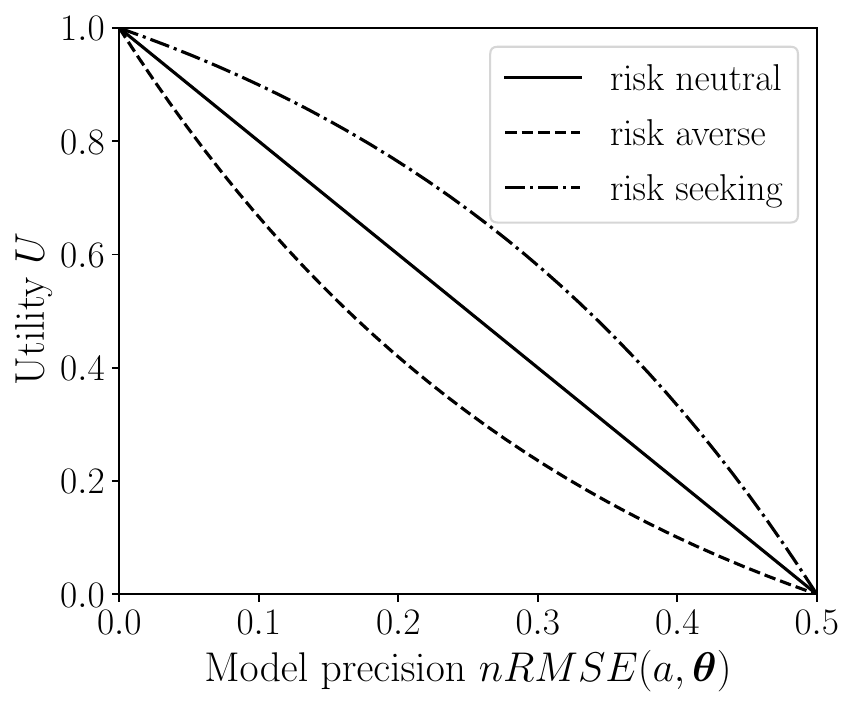}
		\caption{Utility as a function of the model precision attribute. }
		\label{fig:utility_nRMSE}
	\end{figure}

	As discussed above, the model precision, in terms of recovering a target QoI, is one attribute of the DMUU problem for model selection. In this section, the target QoI is assumed to be the response amplitude of mass $m$. We quantify model precision based on the normalized Root Mean Squared Error (nRMSE), which is defined below for both models with respect to the $i$-th posterior sample of $\boldsymbol{\theta}_1$ and $\boldsymbol{\theta}_2$:
	\begin{equation}
		\text{model 1:}\,\,\,\,\,\, nRMSE(a_1, \boldsymbol{\theta}_1^{(i)}) = \sqrt{\dfrac{\sum_{j=1}^{n_f}\left(V_1(f_j, \boldsymbol{\theta}_1^{(i)})\cdot\tilde{y}_{a_w,j} - \tilde{y}_{a_z,j}\right)^2}{\sum_{j=1}^{n_f}(\tilde{y}_{a_z,j})^2}},
		\label{eq:nRMSE_1DOF}
	\end{equation}
	\begin{equation}
		\text{model 2:}\,\,\,\,\,\, nRMSE(a_2, \boldsymbol{\theta}_2^{(i)}) = \sqrt{\dfrac{\sum_{j=1}^{n_f}\left(V_2(f_j, \boldsymbol{\theta}_2^{(i)})\cdot \omega_j^2 \cdot \tilde{y}_{F_h,j} - \tilde{y}_{a_z,j}\right)^2}{\sum_{j=1}^{n_f}(\tilde{y}_{a_z,j})^2}},
		\label{eq:nRMSE_2DOF}
	\end{equation}
	We define the utility function associated with the model precision attribute, i.e., the function $U\big(nRMSE(a, \boldsymbol{\theta})\big)$, which is illustrated in Fig.~\ref{fig:utility_nRMSE}. Different utility curves can be derived according to the decision maker's risk profile \cite{Berger_1985,CHADHA2023108845}. For the investigations that follow, we employ the utility function represented by the solid black curve in Fig.~\ref{fig:utility_nRMSE}, which corresponds to a risk neutral decision maker.

	Let us assume that model precision is the only attribute in the DMUU problem for model selection. The decision concerns which model to use for better simulating (in terms of model precision) the experimentally realized dynamical system of Fig.~\ref{subfig:experiment}. For this decision-making task, we rely on Equation~\ref{eq:MEU}. The expected utility of the decision $a_1$ to employ model 1 is approximated using $n_s$ posterior samples from $\pi_{\mathrm{pos}}(\boldsymbol{\theta}_1|\tilde{\boldsymbol{y}}_{a_z})$ as:
	\begin{equation}	\mathbb{E}_{\boldsymbol{\theta}_1|\tilde{\boldsymbol{y}}_{a_z}}[U(a_1,\boldsymbol{\theta}_1)] \approx \dfrac{1}{n_s}\sum_{i=1}^{n_s} U\big(nRMSE(a_1, \boldsymbol{\theta}_1^{(i)})\big),
	\label{eq:expected_utility_model_1}
	\end{equation}
	while for the decision $a_2$ to employ model 2, using $n_s$ posterior samples from $\pi_{\mathrm{pos}}(\boldsymbol{\theta}_2|\tilde{\boldsymbol{y}}_{a_z})$, we get the approximation:
	\begin{equation}	\mathbb{E}_{\boldsymbol{\theta}_2|\tilde{\boldsymbol{y}}_{a_z}}[U(a_2,\boldsymbol{\theta}_2)] \approx \dfrac{1}{n_s}\sum_{i=1}^{n_s} U\big(nRMSE(a_2, \boldsymbol{\theta}_2^{(i)})\big).
	\label{eq:expected_utility_model_2}
	\end{equation}
	The evaluation of Equations~\ref{eq:expected_utility_model_1} and~\ref{eq:expected_utility_model_2} results in \smash{$\mathbb{E}_{\boldsymbol{\theta}_1|\tilde{\boldsymbol{y}}_{a_z}}[U(a_1,\boldsymbol{\theta}_1)]\approx0.852$ and $\mathbb{E}_{\boldsymbol{\theta}_2|\tilde{\boldsymbol{y}}_{a_z}}[U(a_2,\boldsymbol{\theta}_2)]\approx0.684$}. Therefore, in function of the above-defined model precision attribute, the optimal decision is $a_1$, i.e., to select model 1, as it leads to the larger expected utility. 
	
	A second attribute should in principle be included in this decision problem, namely the computational cost of each model. The cost of the one-mass oscillator model 1 can be reasonably assumed to be smaller than the cost of the two-mass oscillator model 2. Thus, including this cost as an additional attribute would not change the optimal decision, and would instead reinforce the decision $a_1$ to select model 1. In general, when more than one attributes are considered in a decision problem, one needs to refer to multi-attribute utility theory \cite{keeney_raiffa_1993}. 

	\subsection*{Model selection with respect to the frequency content characterization}

\begin{figure}
		\centering
		\includegraphics[width=.40\linewidth]{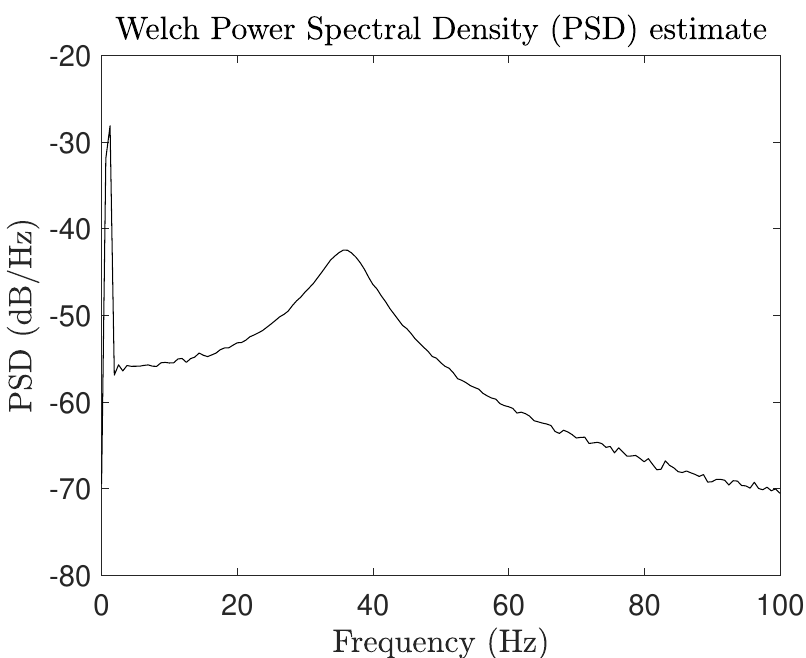}
		\caption{Estimate of the PSD of the input acceleration signal $\ddot{z}(t)$ using Welch's method.}
		\label{fig:PSD}
	\end{figure}

	Let us now assume that the goal is to select the model that best characterizes the frequency content of the realized dynamical system of Fig.~\ref{subfig:experiment}, based on the output-only measurement of the acceleration of the oscillating mass $m$. Hence we now assume that the measurement vector is $\mathbf{z}=[\ddot{z}(t)]$. We estimate the Power Spectral Density (PSD) of the input acceleration signal $\ddot{z}(t)$ using Welch's method \cite{Stoica_2005}. Fig.~\ref{fig:PSD} plots this PSD estimate, which reveals two frequency peaks captured in the estimate, one corresponding to the first eigenfrequency of the frame mass and the other to the first eigenfrequency of the oscillating mass. For some tasks, capturing both frequency peaks is crucial. The one-mass oscillator model 1 of Fig.~\ref{subfig:1DOF} corresponds to a single resonant peak and is thus not sufficient in characterizing the frequency content of the investigated dynamical system. To this end, the two mass-oscillator model 2 is the better model, as it is able to capture both peaks. Thus, in this section we qualitatively point out that, by changing the target QoI, the model selection outcome may differ.
		
	\section{CONCLUSION}
	This paper presents a Bayesian decision-theoretic approach to the problem of selecting a model for simulating the behavior of engineered systems, conditional on the availability of experimental data. The presented approach is numerically investigated on the IMAC-MVUQ Round-Robin Challenge. Specifically, we investigate the problem of deciding to select among two candidate models, a one-mass (model 1) and a two-mass (model 2) oscillator model, the one that most precisely recovers target Quantities of Interest (QoI) derived from experimental monitoring data, while simultaneously taking into account the models' computational cost. The experimental monitoring data are obtained via the realization of the investigated dynamical system as an experimental test environment. When the response amplitude of the oscillating mass is defined as the target QoI, and assuming that measurements of the input excitation are accordingly available for both models, the optimal decision is to select the simple one-mass oscillator model, as it maximizes the expected utility. Conversely, model 1 qualitatively proves insufficient compared to model 2, when we define as target QoI the characterization of the realized dynamical system's frequency content.
	
\bibliography{mybibfile}
	
\end{document}